# Breaking the Reptation Trap: Escape Dynamics of Semi-Flexible Polymers in Crowded Networks

Authors: Ahmad Reza Motezakker[1,2]*, Andrés Córdoba[3], Tomas Rosén[2,4], Fredrik Lundell[1], L. Daniel Söderberg[2,4]*

**Affiliations:**

[1]Department of Engineering Mechanics, KTH Royal Institute of Technology; Stockholm, SE-10044, Sweden

[2]Wallenberg Wood Science Center, KTH Royal Institute of Technology; Stockholm, SE-10044, Sweden

[3]Pritzker School of Molecular Engineering, University of Chicago; Chicago, Illinois, 60637, USA

[4]Fiber and Polymer Technology Department, KTH Royal Institute of Technology; Stockholm, SE-100 44, Sweden

*Corresponding authors. Emails: dansod@kth.se , armot@kth.se

**Abstract:**

Semi-flexible polymers in crowded environments exhibit complex dynamics that play a crucial role in various biological and material design processes. Based on the classic reptation theory, it is generally believed that semiflexible polymers are trapped within static confinements. Here we demonstrate that semi-flexible polymers are indeed trapped in short-lived kinetic cages. We have developed a novel scaling law for rotational diffusion through the examination of the polymer network at the level of an individual nanofiber and quantify the onset of entanglement and entanglement itself by introducing an effective tube diameter. We also show that understanding these dynamics is critical for interpreting the macroscopic behavior of such systems, as evidenced by our micro-rheology analysis. The insight from our study provides a rectified microscopic understanding of the dynamics of semi-flexible polymers in complex environments, yielding new principles to better understand the macroscopic behavior of crowded systems, from biophysics to materials science.

**Main Text:**

Knowledge regarding the dynamics of semi-flexible nanoscale filaments has the potential to (i) increase our understanding of cellular processes such as cell division, cell motility, and intracellular transport and (ii) develop new materials with enhanced mechanical properties, as well as for designing and optimizing nanotechnologies such as nanorobots and nanosensors (1). One example, without any loss of generality, is cellulose nanofibers (CNF), which can be obtained from plants or synthesized by microorganisms (2, 3). CNF's low density, high strength, high stiffness, and chemically adjustable surface are just a few reasons why it's such a promising nanoscale building block for developing high-performance materials. Inspired by significant architectural elements in nature, networks of semi-flexible CNFs are used to provide structural integrity, resistance to deformation, a natural porous scaffold for transporting substances, and chemical functionality for the final bio-based material (4, 5). Given its abundance and processability, CNF is a facile model system that allows facile processing and characterization, similar to using carbon nanotubes.

In condensed matter physics, analyzing the dynamics of a highly concentrated (entangled) dispersion of semi-flexible nanofibers is an interesting and challenging task. The entanglement of a nanofiber leads to a wide range of unexpected characteristics due to the interactions between the chains. At long times, these substances act like viscous liquids, whereas at short times, they are more like rubber. The reptation model of de Gennes (6), Doi, and Edwards (7, 8) is the most well-known theoretical model for a highly concentrated polymer solution. The central concept of the theory is that an entangled polymer chain crawls in a snakelike manner inside a tube. Edwards initially developed the idea of a tube since, due to topological restrictions imposed by its surroundings, a chain is hindered from experiencing lateral deformations. Because of this, the chain acts as if it were confined in a narrow tube that precisely conforms to its shape. Years later, experimental observations verified the idea, and it became evident that the theory had correctly predicted some properties of flexible entangled polymer solutions (9, 10).

Per definition, the persistence length quantifies the bending stiffness of a polymer by relating it to the thermal (Brownian) forcing, i.e. $L_p = \kappa/k_B T$, where $\kappa$ is the bending stiffness, $k_B$ is the Boltzmann constant, and $T$ is the absolute temperature. Thus, a low persistence length indicates wormlike dynamics and a high persistence length indicates a fiber or rod-like behavior.
It was proposed by Odijk (11) that even slight flexibility increases the rotational diffusion of fibers, $1/D_r \propto L^2 L_p$, which contradicts Doi's argument (12) that rotational diffusion is independent of fiber stiffness, $1/D_r \propto L^7 \xi_p^{-4}$, where $\xi_p$ is the mean distance between the fibers. Odijk's prediction was later supported by a thorough experimental investigation into the Brownian motion of carbon nanotubes in a porous static agarose gel network (13) with controlled confinement. However, there are two conceptual perspectives to consider, given the results presented in this work. Firstly, in a realistic entangled system, all fibers experience confinement while undergoing thermal motion in a dynamic matrix instead of a static one. Secondly, the a priori understanding of the agarose concentration at which confinement is missing. Although Kremer verified the reptation motion in entangled systems using simulations (14), scientists like Skolnick and Fixman contend that reptation is not the predominant mechanism in such systems (15, 16). Taking all into account, the reptation model assumes that each chain is slithering in a stationary timeless tube (or at least until happening of pure reptation) and challenges the importance of interchain entanglements. The main goal of our computational experiments is to quantify the whole entanglement problem and link polymer dynamics to the rheological behavior of biopolymer-based materials at larger scales.

To elucidate the dominating mechanisms in entangled systems, we have applied coarse-grained molecular dynamic (MD) simulations using the ESPResSo software package (17), which allows the coupling to hydrodynamic interactions using the Lattice Boltzmann Method (LBM). The details of the modeling can be found in the supporting information. The simulated nanofibers were constructed using a bead-spring model, where the only interaction between the fibers is repulsion. Each simulated case represents a dispersion of semi-flexible fibers having a length $L$, width $b$, and persistence length $L_p$, resulting in a mean distance between the fibers of $\xi_p \left( = \sqrt{\frac{3}{\nu L}} \right)$, where $\nu$ is the number of chains per volume. Using previously reported values for high-charge TEMPO-oxidized CNF (18), We determined values for fiber properties such as Young's modulus (between 4.5 and 18.4 GPa) and fiber dimensions such as aspect ratio.

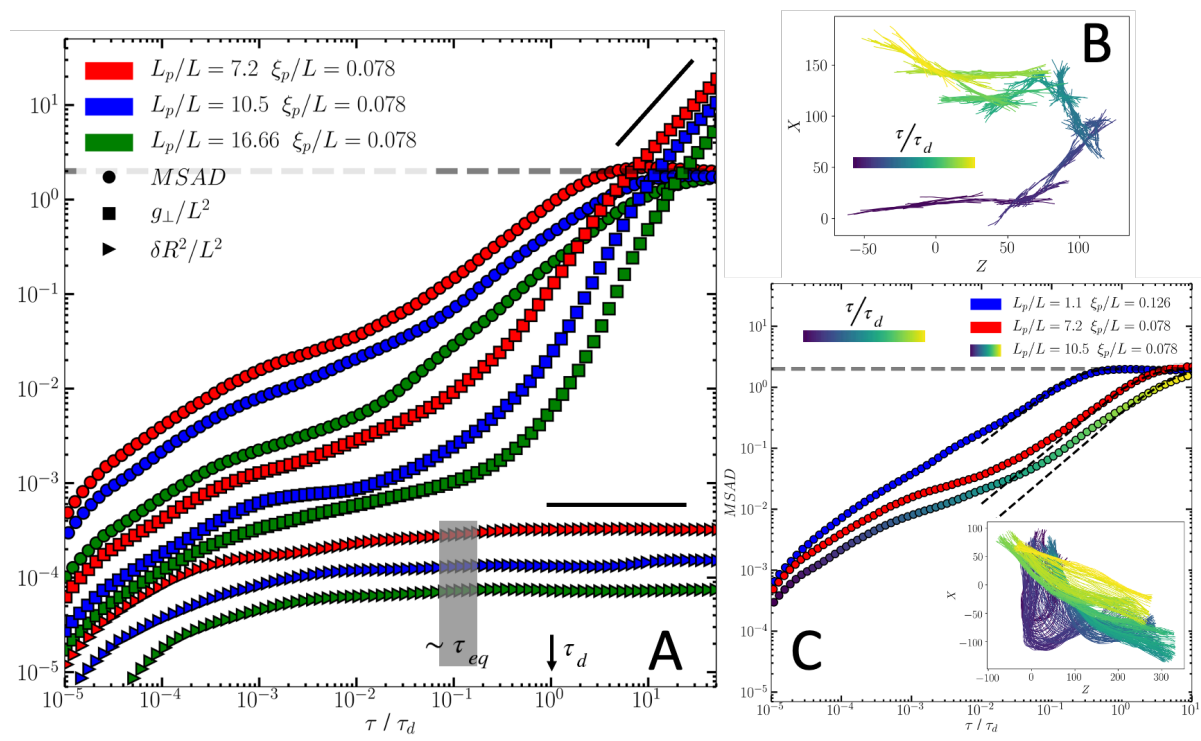

**Figure. 1**. **Dynamics of fibers as a function of stiffness.** (**A**) Projected length fluctuations $\delta R^2$, orthogonal mean squared displacement of centered monomer of fibers $g_\perp$, and mean squared angular displacement of the normalized end-to-end vector $MSAD$ for three $L_P/L$ with a fixed mesh size. (B) Projection of the 3-D motion of a nanofiber onto the x-z plane. (**C**) Comparison of MSAD of entangled and dilute cases.

To characterize the nanofiber dynamics and system relaxation, we have chosen to work with three key non-dimensional parameters. Mean squared displacement of the end-to-end distance (projected length), $\delta R^2(t)$, provides insights into the bending dynamics of individual fibers in the network. Rotational diffusion is characterized using the mean squared angular displacement of the normalized end-to-end vector, $MSAD$. Finally, to provide an understanding of the effect of confinement, the normal mean squared displacement of the mid-point monomer of the fibers, $g_\perp(t)$, will be discussed. In Fig. 1A, we can see the evolution of these parameters for three distinct cases. In a dilute dispersion, where the nanofibers diffuse freely, the projected length fluctuations typically increase with time as $t^{3/4}$, until reaching a plateau value $L^4/45 L_p^2$ at equilibration time $\tau_{eq}$. Crowded systems exhibit similar behavior, but we find a transitional regime before the plateau caused by confinement induced by surrounding fibers, where the initiation correlates to when fibers start experiencing the entanglement. This is referred to as the entanglement time ($\tau_e$). At the end of the transitional regime, when the bending dynamics of the nanofibers $\delta R^2(t)$ have equilibrated the confinement is no longer felt, and the system has reached the equilibrium time $\tau_{eq}$. The equilibrium time appears to be more sensitive to the concentration of the system than it is to the bending stiffness, see supplementary materials, where it is shown that the time required for internal relaxation grows as the crowdedness of the system increases (smaller $\xi_p/L$). Interestingly, the results also show that $\tau_{eq}$ is substantially smaller than the time $\tau_d$ it takes a nanofiber to diffuse its length $L$. It should be noted that this contradicts the fundamental assumptions for the reptation theory, which postulates permanent or long-lasting confinement of the polymer.

By studying $g_\perp(t)$, it is possible to estimate the direct effects of confinement. Similarly to $\delta R^2(t)$, it initially increases with time as $t^{3/4}$ (19), followed by a transition that seems related to the entanglement time, i.e., when the bending fluctuations of the fibers encounter confinement from their surroundings. The transition between these two states serves as a point of reference for calculating hypothetical tube diameter, $d$, (see supplementary information). Our results indicate a relation between the tube diameter and the mesh size of the system as $d \sim \xi_p^{1.2} L_p^{-0.2}$. Given that most studies assume that the tube diameter is identical to the mesh size, this is a surprising finding. After escaping their initial confinement, $g_\perp(t)$ indicates a super-diffusive behavior, presumably traveling a curved route, a motion that couples translation and rotation of the nanofibers (20, 21). Finally, $g_\perp(t)$ shows linear diffusion behavior upon reaching the longest relaxation time, $\tau_r$.

Furthermore, $MSAD$ provides insight regarding $\tau_r$, which is proportional to the inverse of rotational diffusion. Considering $u(t)$ as the unit vector of end-to-end distance of nanofiber, we can write the unit vector correlation function as
$$\langle u(t+\tau) \cdot u(t) \rangle \sim \langle \cos(\theta) \rangle \sim \exp(-2D_r t),$$
where $\theta$ is angular displacement and $2D_r$ is proportional to decay time which is $1/\tau_r$. So, we can write $MSAD$ as
$$\langle (u(t+\tau) - u(t))^2 \rangle = 2(1 - \exp(-2D_r t)).$$
At short times when both $D_r t$ and $\theta$ are small, we can consider $\cos(\theta) \sim 1 - \theta^2/2$, resulting in $MSAD \sim 4D_r t$. However, at longer times, $t \to \infty$, the initial and final direction of the unit vector will be uncorrelated, $\langle u(t+\tau) \cdot u(t) \rangle = 0$, and $MSAD$ will consequently saturate to 2. The same $MSAD$ behavior can be observed regarding the diffusion of free semi-flexible fibers in a dilute dispersion (Fig. 1.B). As expected, confinement in the crowded systems introduces a separation between the short- and long-term behavior of $MSAD$. On the long term, both $MSAD$ and trajectory analysis indicate that fibers rotate like rigid rods (Fig. 1.C). Therefore, we can estimate $D_r$ by fitting the long-term *MSAD* data to $2(1 - exp(-2D_r t))$, which is predicted for rigid rods(7).

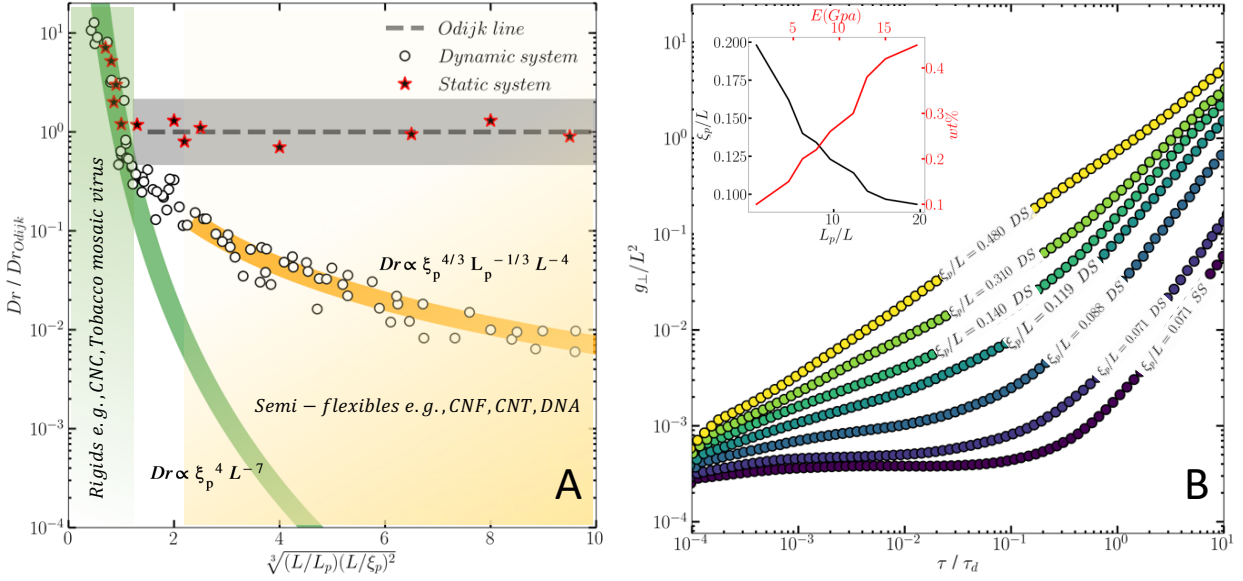

**Figure. 2. Diffusivity of nanofibers.** (A) Normalized rotational diffusivity of nanofibers in a dynamic system with various $L_P/L$ and $\xi_p/L$ versus the length normalized with the deflection length . The dashed grey line represents Odijk's prediction, and the green line Doi's prediction. The grey shading represents Fakhri's (13) experimental. (B) The gradual decrease in the concentration of nanofiber dispersion for case of $L_P/L = 3.2$, illustrating the transition from a robust tubing effect to one that is weak or not observable in $g_\perp/L^2$. The inset demonstrates a correlation between the nanofiber stiffness and the onset mesh size (concentration) in which tubing first appear. The tubing effect may occur at a far lower network concentration if nanofibers are more flexible.

To facilitate comparison, we present $D_r$ in Fig. 2.A in the same manner as in Fakhri et al.(13). The vertical axis represents the ratio of $D_r$ to Odijk's model prediction, while the horizontal axis is $\mathscr{L} = \sqrt[3]{(L/L_p)(L/\xi_p)^2}$, which is considered here as length normalized by the deflection length of the thermal undulation of the fibres. In principle, $\mathscr{L}$ is proportional to the inverse of the stiffness of the nanofibres. Given the obtained curves, we define various $\mathscr{L}$ ranges for data analysis. For $\mathscr{L} < 1$, it is evident that $D_r$ agrees with the Doi's and Odijk's predictions as $D_r \propto \xi_p^4 L^{-7}$, suggesting that fibers with a length of $\mathscr{L} \leq 1$ act like rigid rods and that flexibility does not affect rotational diffusion. Within this limit, the largest possible orthogonal fluctuations of bending modes, $L^3/L_p$, are smaller than $\xi^2$, and Doi and Edwards' mechanism of reptation-rotation predicts the motion of the fibers. However, for $\mathscr{L} \geq 2.3$, flexibility does have a significant impact, where $D_r \propto \xi_p^{4/3} L_P^{-1/3} L^{-4}$; indicating that terminal relaxation occurs faster than Doi's prediction, $D_r \propto \xi_p^4 L^{-7}$, but slower than Odijk's $D_r \propto L^{-2} L_p^{-1}$. The shaded region in Fig. 2A represents Fakhri results, consistent with Odijk's and the simulations by Lang & Frey (22). To better understand the effect of the tube assumption and the discrepancy between our result and Odijk's prediction, we calculated several cases with fibers moving in a static entangled system, in contrast to the previous cases in which all fibers are dynamic. These results lie within the shaded area, thus confirming Odijk's scaling of $D_r \propto L^{-2} L_p^{-1}$. In Fig 2.B., we assess the orthogonal fluctuation of fibers to develop an understanding of the mechanism of confinement for the case with $\xi_p/L$ and $L_p/L$. A closer look at the case of $\xi_p/L = 0.071$ in a static system reveals that $g_\perp(t)$ is relatively stable for a while, indicating that the fiber is confined inside the tube in a static environment until it exits the tube, rotates, and enters a new tube which confirms the classic reptation theory and other simulations with stationary barriers(23, 24). In the identical case, but with all nanofibers moving in a dynamic system, there are similarities, where $g_\perp(t)$ initially is constant, followed by an increase with time, having a slope similar to the static environment. However, the transition point occurs much earlier. This illustrates that the static tube assumption fails owing to the reconfiguration of the surrounding fiber system, and this process is similar to Fixman's Kinetic cage hypothesis (25)(See supplementary materials). The validity of the tube assumption is thus coupled to a clear separation of time scales between the particle concerned and its surroundings. As discussed by Lang & Frey (22), the conditions for which the tube assumption is valid have largely been ignored and are thus still up for discussion. In the inset in Fig. 2.B, the relation between the onset condition of tubing and the fiber stiffness is shown. This graph demonstrates that the tubing effects arise at lower network concentrations for more flexible fibers. It is suggested that this originates in the fact that more flexible fibers show larger orthogonal fluctuations, resulting in, on average, narrower instantaneous pathways for adjacent fibers and, ultimately, the formation of confinements with lower concentrations.

The simulations not only allow the identification of distinct regimes and possible mechanisms related to the dynamics of entangled CNF dispersions, but also allows the examination of the viscoelastic behavior of such systems. This was accomplished by using the passive microrheology technique, allowing the quantification of the sample's rheological characteristics by the introduction of a spherical particle (26) (see supplementary information).
The rheology of the considered dispersions can be found in Fig. 3.A. At low frequencies, $\omega \lesssim 3 \times 10^2 \, rad/s$, $G'(\omega)$ follows a $\sim \omega^2$ scaling, which is expected in the terminal zone of the storage modulus of a viscoelastic fluid. In the terminal zone, all the mechanical stress stored in the viscoelastic fluid has relaxed. The terminal zone is also observed in Fig. 3.B in $G''(\omega)$, where the scaling is $\sim \omega^1$, which is the characteristic scaling in the loss modulus of viscoelastic materials. This scaling in $G''$ reflects the purely viscous response expected in the terminal zone. Both $G'$ and $G''$ exhibit a $\sim \omega^{2/3}$ region at intermediate frequencies $3 \times 10^2 \, rad/s \lesssim \omega \lesssim 10^4 \, rad/s$. This $2/3$ scaling exponent has been observed for networks of actin filaments and is typical of Zimm chains that interact hydrodynamically(27). At high frequencies, $\omega \gtrsim 10^4 \, rad/s$, $G'$ exhibits a transition towards what appears to be a high-frequency plateau. This tendency of $G'$ to exhibit a transition towards a high-frequency plateau is more pronounced for networks with smaller mesh sizes. At higher frequencies, $\omega \gtrsim 10^4 \, rad/s$, $G''$ goes as $\sim \omega^1$, which indicates that for the networks with larger mesh sizes, the purely viscous response of the solvent dominates at high frequencies. For the dispersions with smaller mesh sizes, this scaling is only observed at $\omega \gtrsim 10^5 \, rad/s$, indicating that for the more concentrated networks, the viscoelastic response becomes dominated by the solvent only at frequencies higher than $10^5 \, rad/s$. It is also important to note that for large mesh sizes, $\xi_p/L \gtrsim 0.24$, $G''$ is larger than $G'$ in the whole frequency range studied here.

At around $\xi_p/L = 0.24$ a high frequency crossover between $G'$ and $G''$ appears. For mesh sizes smaller than $\xi_p/L \approx 0.24$, $G'$ becomes larger than $G''$ for $\omega \gtrsim 7 \times 10^3\ rad/s$. As a rule of thumb for defining a transition threshold from flowing to non-flowing behavior of high-charge CNF systems, the mesh size of $\xi_p/L \approx 0.24$ may be used, which corresponds to concentration of 0.06 wt % which is in agreement with previous experiments (28, 29).

Fig. 3.C shows the zero-shear viscosity of the cellulose dispersion with different mesh sizes, where the viscosity increases with decreasing mesh size. We have compared our numerical findings with experimental results (30), notwithstanding the difficulty of determining zero shear viscosity using experimental techniques. As can be seen, they agree when some assumptions are considered (see supplementary materials). For mesh sizes larger than about $\xi_p/L \approx 0.24$ the viscosity goes as $(\xi_p/L)^{-4}$. At $\xi_p/L \approx 0.24$ there is a clear transition on how the zero-shear viscosity depends on the mesh size. For $\xi_p/L \lesssim 0.24$ the viscosity increases slower with decreasing mesh size, $\eta \sim (\xi_p/L)^{-2}$. Note that the mesh size at which this change in the behavior of the viscosity occurs is the same mesh size at which the crossover between $G'$ and $G''$ starts to appear.

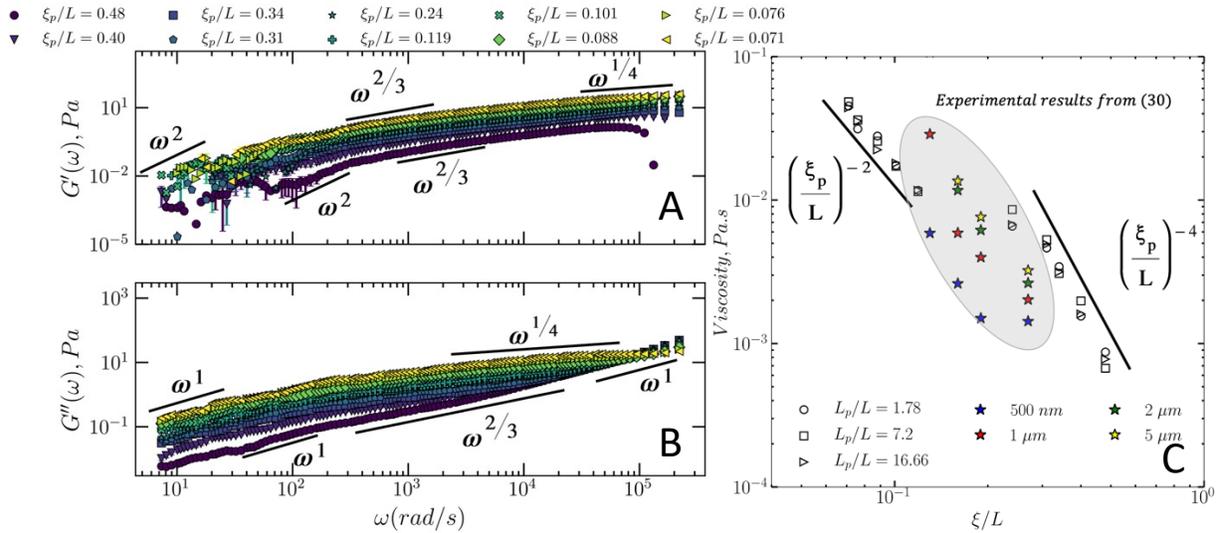

**Figure. 3. Microrheology of the dispersion and comparison to experiments. A)** Storage modulus, $G'(\omega)$, and **B)** Loss modulus, $G''(\omega)$, measured with passive microrheology of CNF dispersion with $L_p/L = 7.2$ for different mesh sizes, $\xi_p/L$. **C)** Zero shear viscosity of a cellulose dispersion with different mesh sizes, $\xi_p/L$ measured passive microrheology. Colored markers represent experimental data (30).

Although a thorough understanding of cellulose nanofiber (CNFs) networks has been proven essential for developing processing protocols for high-performance applications (31), knowledge has, as stated in the introduction, application to other engineering and biological systems with nanofibers, e.g., carbon nanotubes, metallic nanowires, or protein fibrils. Based on our findings, we propose a novel scaling law for the rotational diffusion of semi-flexible nanofibres in entangled systems, and we demonstrate the coupling between the nanofibers' rotating and translational motions. To supplement the study, we have also conducted numerical microrheology experiments, which illustrate a way of using simulations together with microrheology as a tool to elucidate the nanoscale dynamics.

**Methods**

Methods and any associated references are available in the supplementary material.

**Data and availability:** All data are available in the main text or the supplementary materials.

**Acknowledgments:** The authors acknowledge the discussions and input to the work and the manuscript provided by Prof. Theo Odijk, Leiden University.

**Author contributions:**

Conceptualization: ARM, LDS, TR, FL.

Methodology: ARM, LDS, AC, TR.

Investigation: ARM, LDS, AC, TR.

Funding acquisition: LDS.

Project administration: LDS.

Supervision: LDS, TR and TR.

Writing – original draft: ARM, LDS, AC.

Writing – review & editing:  ARM, LDS, AC, TR and FL.

**Competing interests:** Authors declare that they have no competing interests.

**Funding:**

Knut and Alice Wallenberg Foundation (KAW) through the Wallenberg Wood Science Center (ARM, FL, LDS).

The Swedish Research Council grant 2018-06469 (TR)

# Title: Breaking the Reptation Trap: Escape Dynamics of Semi-Flexible Polymers in Crowded Networks

**Authors:** Ahmad Reza Motezakker[1,2]\*, Andrés Córdoba[3], Tomas Rosén[2,4], Fredrik Lundell[1], L. Daniel Söderberg[2,4]\*

**Affiliations:**

[1]Department of Engineering Mechanics, KTH Royal Institute of Technology; Stockholm, SE-10044, Sweden

[2]Wallenberg Wood Science Center, KTH Royal Institute of Technology; Stockholm, SE-10044, Sweden

[3]Pritzker School of Molecular Engineering, University of Chicago; Chicago, Illinois, 60637, USA

[4]Fiber and Polymer Technology Department, KTH Royal Institute of Technology; Stockholm, SE-100 44, Sweden

\*Corresponding authors. Emails: dansod@kth.se , armot@kth.se

**Modeling**

This section describes a coarse-grained model for simulating semi-flexible polymer systems. Polymers and the fluid (liquid solvent) are the two basic elements of such systems that must be accounted for in the modeling procedure. The ESPResSo software package was used for all simulations. A bead-spring chain model was used to represent the polymer, with a Weeks-Chandler-Andersen (WCA) potential set up between the beads as below

$$U_{WCA}(r) = \begin{cases} 4\varepsilon\left[\left(\frac{\sigma}{r}\right)^{12} - \left(\frac{\sigma}{r}\right)^{6}\right] + \varepsilon, & \text{if } r < 2^{\frac{1}{6}}\sigma, \\ 0, & \text{else}, \end{cases}$$

where $\sigma$ is the diameter of the beads and $\varepsilon$ represents the energetic pre-factor with Boltzmann constant $k_B$ and temperature $T$. An attractive potential is used to explicitly resolve each polymer chain utilizing MD beads that are coupled to one another; in this case, we use a standard harmonic interaction in the form of

$$\varphi_{harm} = \frac{1}{2}k(r - r_0)^2,$$

where $r$ is the distance between two adjacent MD beads in a polymer chain, $r_0$ is the equilibrium distance of the potential, and $k$ is the spring constant that defines the bond stiffness. The angle potential is used to directly tune the nanofiber bending rigidity as follows

$$V(\varphi) = K(1 - \cos(\phi - \phi_0)),$$

where K is the bending constant and phi0 is the equilibrium bond angle in radians ranging from 0 to $\pi$.

We coupled the beads to a thermalized lattice-Boltzmann (LB) fluid to include hydrodynamic interactions. We were able to successfully couple the LB fluid and the MD beads relying on the point-coupling strategy provided in (1), which coupling force is as follows

$$\vec{F} = -\gamma(\vec{u}_{fluid} - \vec{u}_{bead}),$$

where $\gamma$ is the friction parameter, $\vec{u}_{fluid}$ is the fluid velocity, and $\vec{u}_{bead}$ is velocity of the bead. Therefore, the equation of motion is the same as the Langevin equation, except that it considers the bead's relative velocity to the fluid as follows

$$m_i \dot{v}_i(t) = f_i - \gamma(v_i(t) - u(x_i(t), t)) + \sqrt{2\gamma k_B T}\, \eta_i(t)$$

Here, $f_i$ represents all deterministic forces resulting from interactions and $\eta$ a random thermal force. The friction term allows for dissipation in the surrounding fluid, whereas the random force simulates collisions of the particle with solvent molecules at temperature $T$ and meets the condition

$$\langle \eta(t) \rangle = 0, \text{ and } \langle \eta_i^\alpha(t)\, \eta_j^\beta(t') \rangle = \delta_{\alpha\beta}\, \delta_{ij}\, \delta(t - t'),$$

where $\langle . \rangle$ denotes the ensemble average, $\alpha, \beta$ are spatial coordinates, and $\delta$ is the Kronecker delta. The LB fluid is discretized using a D3Q19 geometry with a grid spacing of $a = \sigma$.

ESPResSo employs the notion of simulation units, which enables the user to freely choose the system's mass, length, and energy scales, which together define all other parameters. The nanofiber width, which we assumed to be 3 nm, is the shortest length scale in the system; hence, we define length scale [x]=3 nm. As simulations are conducted at $T = 25\,°C$, we employ $T \approx 300\,K$ in our model and provide the energy scale [E]= $k_B.300\,K$. The unit mass of a single bead was set to [m]=1.

In implicit-solvent simulations, it is not possible to simply calculate the time scale by translating mass, energy, and length units to time units in a simulation as

$$[t]=[x]\cdot\sqrt{\frac{[m]}{[E]}},$$

because the implicit solvent (LB fluid) determines the viscosity of the solvent, which does not enter the time unit calculation. Therefore, we determine the time unit by using the rotational diffusion coefficient of a stiff rod of the same length as the shortest nanofiber in our system. According to Doi's prediction for a rigid rod

$$D_r \approx \frac{3 k_B T \left(2\ln\left(\frac{2L}{b}\right) - 1\right)}{16\pi\mu\left(\frac{L}{2}\right)^3},$$

where L is the rod length, b is the width, and $\mu$ is the dynamic viscosity of the solvent, which is water in our case $\mu_w \approx 8.9\times 10^{-4}\,Pa.s$. $D_r^{-1}$ has the unit of time, and our simulations are on the order of $[t] \approx 50\,ns$. An outline of the simulation unit system's parameters for a few representative system parameters is presented in Table. S1.

**System setup**

First, a cubic periodic box is defined for the simulations. To avoid any boundary effect on the nanofibers' dynamics, we ensure that the box size exceeds the nanofiber length. for the presented simulations, we have kept the minimum size of $1.3\,L$ for the box size, where $L$ is the nanofiber length. The following step is to define polymer chains using the so-called bead spring model. In order to mimic high-charged tempo cellulose nanofiber (CNF), we choose nanofiber lengths that are within the reported length distribution of this particular CNF type (50<AR<300). Assuming that the cross section of the fibers is circular, the area moment of inertia can be determined as $I = \pi\sigma^4/64$, and area as $A = \pi\sigma^2/4$. The bead-spring model is able to separate the Young's modulus and the bending rigidity of the coarse-grained polymer since they are driven by bond and angle terms, respectively. The bond potential is given by

$$\varphi_{harm} = \frac{k}{2}(r - r_0)^2,$$

and the semiflexible nanofiber has a modulus proportional to its spring constant given by $k = AE/r_0$, where $E$ is Young's modulus and $A$ is the cross-sectional area of the nanofiber. To have direct control over the bending rigidity, the angle potential is chosen accordingly. A polymer chain's bending energy is given by

$$E_B = \frac{\kappa}{2}\int_0^l C^2 ds = \frac{\kappa}{2}\int_o^l \left|\frac{\delta T}{\delta s}\right|^2 ds = \frac{\kappa}{2} r_0 \sum_i \left(\frac{\Delta T}{r_0}\right)^2$$

$$= \frac{\kappa}{r_0}\sum_i (1-\cos(\pi-\theta)),\ 0<\theta<\pi$$

where $\kappa$ is bending rigidity, $C$ is local curvature, $l$ is the contour length, and $T$ is unit tangent vector. Here the angel potential is chosen as $V_{bending}(\theta) = \kappa/r_0\,(1-\cos(\pi-\theta))$ and the relationship between Young's modulus and bending rigidity may be expressed as $\kappa = EI/r_0$. It is possible to test this assumption by juxtaposing the computed bending rigidity of coarse-grained semiflexible polymer chains from simulations with the bending rigidity specified in the angle potential. I brief, the bending rigidity of a coarse-grained semiflexible polymer chain may be determined by clamping one end and monitoring the thermally-induced variation of the other end as

$$\kappa^* = \frac{k_B T L^3}{3\langle \delta u(L)^2 \rangle}$$

where $L$ is the length of the semi-flexible polymer chain and $\delta u(L) = u(L) - \langle u(L)\rangle$ is the deviation of the free end from its average position(2). Although there is a natural high variance in measured bending rigidity by applying thermally generated contour fluctuations, the assigned bending rigidity and the observed bending rigidity are in agreement with an acceptable error (see Fig S.1). This allows us to readily determine the nanofibers' persistence length $L_p = \kappa/k_B T$. Finally, we model the CNFs using the Young's modulus given in the literature, which is in the range of 4.5–18.4 GPa for CNF (3).

When the nanofibers are built, we scatter them randomly within the box. Thus, it is conceivable for particles to overlap, resulting in a very strong force of repulsion. In this scenario, numerical instability prevents us from integrating the equation of motion. To remove the overlapping, we place constraints on the maximum force that may act between two particles, integrate the equation of motion, and gradually increase the force limit between two particles. Next, we let the chains stabilize by integrating the equation of motion for a while. In the next step, we couple the beads to the fluid by defining the LB fluid and its associated parameters and then applying the LB thermostat to define the friction coefficient. Here, we integrate the equation of motion to warm up the system, including the nanofibers and the LB fluid. The size of the system, such as the number of particles and the length of the polymer, greatly affects the number of steps for the warm-up operation. At long last, we've reached the stage where we can begin integrating the system that records trajectories and other observables. A demonstration of nanofiber modeling for the special situation of $L_p/L = 1.5$ and $\xi_p/L = 0.12$ in Fig. S2. In supplementary movie S1, the dynamics of nanofiber simulations are also shown for the same case.

### Observables

Since the position and velocity of each bead are sampled in each time step of the simulations, practically all the system's statistics are accessible. When reporting observables, data must be averaged throughout all chains and measurements taken at different times. This is achieved via the use of ensemble averaging, $\langle . \rangle$.

Each nanofiber chain is composed of $N$ beads with an end-to-end vector connecting the first bead to the final bead as $\boldsymbol{R}(t) = \boldsymbol{r}_N(t) - \boldsymbol{r}_1(t)$. For internal bending relaxation analysis, end-to-end vector fluctuations are used as

$$\delta R^2(\tau) = \langle (|R(t+\tau)| - |R(t)|)^2 \rangle$$

Since the arc length of a polymer remains constant as it fluctuates, the size of the end-to-end vector must be smaller than the actual length of the polymer itself. These fluctuations enable the polymer to reptate along its baseline axis in a semi-dilute environment, when the polymer concentration is high enough that the entanglement length ($L_e$) becomes less than the persistence length ($L_p$). It should be noted, nonetheless, that nanofiber rotation plays no role in $\delta R^2(t)$.

Granek (4) showed that at short times $\delta R^2(t)$ is roughly

$$\delta R^2(t) \simeq const\, L \left(\frac{k_B T}{\kappa}\right)^{5/4} \left(\frac{k_B T}{\mu} t\right)^{3/4}$$

And on the other hand, as $t \to \infty$ end-to-end fluctuations saturate to

$$\delta R^2 = \frac{1}{45}\left(\frac{k_B T}{\kappa}\right)^2 L^4$$

What this implies is that $\delta R^2$ acts like $t^{3/4}$ for short times until internal relaxation time, but eventually saturates to $L^4 / 45 L_p^2$.

It is generally expected that the nanofibers' contour would bend less dramatically in semi-flexible circumstances, where $L_p \gg L$, than in flexible ones, where $L_p \ll L$. Therefore, we examine the orientation of the end-to-end vector to assess the rotational motion of the nanofibers. We calculate mean squared angular displacement (MSAD) for the unit vector of the end-to-end vector of each chain, which is defined as

$$\vec{u}_R(t) = R(t)/|R(t)|$$

and MSAD as

$$MSAD = \delta u_R^2(\tau) = \langle (u_R(t+\tau) - u_R(t))^2 \rangle$$

Using MSAD, one can determine the nanofibers' rotational diffusion, $D_r$, and the longest relaxation time of the system, $\tau_r \propto D_r^{-1}$.

Next, we assess the mean-squared displacement (MSD) of the nanofibers' centered bead (CB) to learn about the parallel and normal diffusion and ensuing tubing mechanism in a crowded environment as follows(5)

$$g_\parallel(\tau) = \langle [(r_{CB}(t+\tau) - r_{CB}(t)) \cdot u_R(t)]^2 \rangle$$
$$g_\perp(\tau) = \langle [(r_{CB}(t+\tau) - r_{CB}(t))]^2 \rangle - g_\parallel(\tau)$$

For times shorter than the system's internal relaxation time, contour fluctuation, $\delta R^2(t)$, is the main mechanism, and as a result, $g_\perp(t)$ behave similarly to $\delta R^2(t)$ as $t^{3/4}$ (4).

**Supplementary text**

A comment on nanofiber motion in a dynamic entangled system

To comprehend the deviation of the results from theories and earlier simulations, we must scrutinize their basic conditions and assumptions. The initial reptation theories (6, 7), for instance, postulate a fixed tube for the polymer, which is only renewed when the polymer reaches a new tube at either end. They believe that polymer chains may diffuse down the tube, but that polymer motion in the orthogonal direction is hampered more than the tube diameter. This hypothesis is based on the fact that the trapped polymer, in order to push the tube, must overcome the drag force associated with tube, which is made up of several surrounding chains. The drag effect in the orthogonal direction is further amplified by the fact that each chain has its own surroundings. Consequently, a stationary tube might be considered in the first place. However, in the real system, all the chains undergo thermal motion, compromising the tube's stability. It is possible for a tube to distort, slide, or even collapse entirely, all without allowing the trapped polymer to reptate its own length. Several constraint release mechanisms, such as tube length fluctuations (8, 9), tube enlargement (10, 11), and convective constraint release (12–14), have been proposed in the literature.

We have studied the dynamics of single nanofibers across a wide range of time scales to evaluate the constraint release mechanism. In Supplementary Movie S2, for instance, we see the dynamics of a single nanofiber for the case of $L_p/L = 7.2$ and $\xi_p/L = 0.098$ (an extra example is shown in Movie S3 for shorter nanofibers). To better comprehend the constraint release mechanism, the nanofiber's dynamic is broken down into multiple stages in Fig. S3. The nanofiber is presented in an entangled environment in the first scene (A). As time passes, it becomes evident that the nanofiber reptates along the tube's length while remaining confined inside the tube (B). The nanofiber will continue to reptate in the next step (C), and at this point, the form (red dashed line) and diameter of the tube will be distinguishable. The fourth phase (D) is the point in time when the tube begins to disassemble, and the constraint is released. The nanofiber breaks out from the tube, as shown by the black arrow, and immediately begins to investigate hitherto unexplored regions of space. In the subsequent stage (E), the nanofiber will continue to freely diffuse while undergoing a turning motion. This motion highly couples the nanofiber's translation with its rotation. This phenomenon can explain the super-diffusive behavior of $g_\perp(t)$ plots for entangled systems. The nanofiber continues its exploration until it is eventually confined in a new tube (F). To the best of our knowledge, the vast majority of research on the constraint release mechanism has not taken into account semi-flexibility. Since the influence of tube length fluctuation and tube enlargement are minor owing to the semi-flexibility of the nanofibers, our simulations indicate that the constraint release mechanism is more similar to the convective constraint release mechanism.

Understanding the mechanism of constraint release requires incorporating Marshal Fixman's notion of kinetic cages (15, 16) and Bitsanis(17, 18) proposed model. Fixman's theory proposes that the entropic cage surrounding a polymer is not static but instead undergoes fluctuations due to the thermal motions of the surrounding polymers. These fluctuations can lead to torque fluctuations, which can break the cage and increase the reorientation dynamics of the polymer. In essence, Fixman's theory suggests that polymer motion in crowded environments is governed by the interplay between entropic confinement and thermal fluctuations, and that these dynamics can be understood by considering the geometry and fluctuations of the entropic cage surrounding the polymer. Bitsanis' model suggests that many-body effects contribute to the static properties much less than direct pair interactions; therefore, the local structure is mostly determined by independent binary interactions. The model suggests that the fluctuating entropic cage can create regions of high and low friction that affect the reptation dynamics of the polymer.

To gain a deeper understanding of the constraint release mechanism, we conducted two sets of simulations involving polymers with infinite persistent lengths (rigid rods). In the first set, a test rod is introduced into a static network of anisotropic rods, while in the second set, all the rods in the network are allowed to move (supplementary Movies S6 and S7, respectively). In the first case, the entropic cage surrounding the rigid rod polymer remained stable and static throughout the simulation. However, in the second case, the surrounding rod polymers were moving and continually bumping into the test rod, leading to fluctuations in the entropic cage surrounding the polymer. This is consistent with Bitsanis' model, which takes into account the dynamic nature of the entropic cage surrounding the polymer. The torque fluctuations observed in the dynamic case are a key mechanism underlying the dynamics of the entropic cage surrounding the polymer. As the surrounding polymers bump into the test rod, they create torques that can break the entropic cage, leading to faster reorientation dynamics of the polymer. Additionally, the second virial coefficient, which describes the strength of the excluded volume interactions between polymer chains, can play an important role in the dynamics of the entropic cage and the overall reptation motion of the polymer. In the dynamic case, the second virial coefficient may influence the density and arrangement of the surrounding polymers, which, in turn, can affect the entropic cage surrounding the polymer and the dynamics of the reptation motion. Movie S8 depicts the empty gaps between the nanofibers with time, thereby providing a better understanding of kinetic cages and how imaginary tubes arise and dissolve.

A comment on experimental results of Fig. 3C

The experimental results (stars) in Fig. 3C are taken from Fig. 7 and Fig. 3 of Nordenström et al. (19). It is essential to note that we have made certain assumptions, such as using the average length given distribution given to calculate $\xi_p/L$. Moreover, we have considered the Stokes-Einstein assumption for the diffusion coefficient $D = \frac{k_B T}{3\pi\eta a}$.

**Microrheology technique**

In the technique called passive microrheology the mean-squared displacement (MSD) of a micron-sized bead embedded in a viscoelastic fluid is used to infer the dynamic modulus, $G^*(\omega)$, of the fluid. To obtain the MSD of the probe bead, the $\gamma$-positions ($\gamma=(x,y,z)$) of the probe bead as a function of time, $r_{b,\gamma}(t)$, is tracked. Where $t=i\Delta t$, $1/\Delta t$ is the sampling frequency of the measurement device, $N$ is the total number of measurements and $i=0,1,2,3,\ldots,N$. The first step in the data analysis requires the calculation of the mean-squared displacement (MSD) of the probe bead,

$$\langle \Delta r^2(\tau) \rangle := \frac{1}{n} \sum_{\gamma} \sum_{i=1}^{n} \Delta r^2_{i,\gamma}(\tau)$$

where

$$\Delta r^2_{i,\gamma}(\tau) := \left[ r_{b,\gamma}(t+\tau) - r_{b,\gamma}(t) \right]^2, n = N - \tau/\Delta t, \tau$$

is the lag time and the sub-index $\gamma := \{x,y,z\}$ indicates the spatial direction of the measurement.

In order to obtain an estimate of the statistical uncertainty in $G^*(\omega)$ obtained from passive microrheology, one needs to start by estimating the statistical error in the MSD. A common omission in the analysis of passive microrheology data is to neglect correlations inherent in the bead position data. These correlations are very important in viscoelastic materials. In a typical viscoelastic fluid, the uncertainty in the MSD of the probe bead can be as high as $20\%$. The highest errors occur at long lag times. At those long times, the uncertainty is underestimated by a factor of about 20 if the correlation in the bead position data is neglected (20). Here we use the Microrheology Uncertainty Calculation Helper (MUnCH) (20, 21) to calculate $\langle \Delta r^2(\tau) \rangle$ and its uncertainty, $\sigma(\langle \Delta r^2(\tau) \rangle)$. By using repeated block transformations, MUnCH can correctly estimate the statistical error of any autocorrelation at any given lag time. Examples of MSDs of the probe bead and their uncertainties in a cellulose solution with different mesh sizes, $\xi_p/L$, are shown in Fig. S.7. Note that the overall magnitude of the MSD of the probe bead decreases with decreasing mesh size of the cellulose solution.

In passive microrheology, the dynamic modulus of the host medium, $G^*(\omega)$, is calculated using the Generalized Stokes-Einstein Relation (GSER). However, applying the original GSER to molecular dynamics simulations with periodic boundary conditions can result in overestimated $G^*(\omega)$ values because of the hydrodynamic interaction between the probe bead and its periodic images. A correction to the GSER has been derived by implementing an analytical solution for Stokes drag on a periodic array of spheres, which allows smaller box sizes to be simulated while still retaining accuracy (22). This relation has been called the hydrodynamic Generalized Stokes-Einstein Relation (HGSER). The HGSER allows the use of smaller box sizes, reducing computational costs by more than an order of magnitude (22)

$$G^*(\omega) = \left( K_S \frac{f_V}{f_M} \right)^{-1} \left( \frac{k_B T}{\pi R i \omega \langle \Delta r^2[\omega] \rangle} \right)$$

Where $R$ is the bead radius, $k_B$ is the Boltzmann constant, $T$ is the temperature, $\omega$ is the radial frequency and

$$\langle\overline{\Delta r^2}[\omega]\rangle := \int_0^\infty \langle\Delta r^2(\tau)\rangle e^{-i\omega\tau} d\tau$$

is the one-sided Fourier transform of the MSD of the probe bead.

The factor

$$\left(K_S \frac{f_V}{f_M}\right)^{-1}$$

accounts for the interactions of the probe bead with its periodic images, where $f_M$ is the mass fraction and $f_V$ the volume fraction of fluid (solvent + cellulose fibers) in the simulation box. For the systems considered in this work $f_M \approx 0.5$ and $f_V \approx 0.95$, the exact values for each particular system are calculated and used in the data analysis of each system. The function $K_S$ as a function of the volume fraction of spheres $\phi = 4\pi/3 (R/L_{box})^3$ is given by (22)

$$K_S^{-1} = 1 - 1.7601\phi^{1/3} + \phi - 1.5593\phi^2 + 3.9799\phi^{8/3} - 3.0734\phi^{10/3}$$

Where $R$ is the radius of the probe bead and $L_{box}$ is in the systems considered here $R = 166.66\,\sigma$, $T = 300\,K$ and $L_{box} = 600\,\sigma$ is the length of the cubic periodic box.

To calculate $\langle\overline{\Delta r^2}[\omega]\rangle$ we use a method that does not require fitting $\langle\Delta r^2(\tau)\rangle$ with an analytic function to perform the one-sided Fourier transform (23). Specifically, we use

$$i\omega\langle\overline{\Delta r^2}[\omega]\rangle \approx (2\pi i)^{-\alpha(\omega)} \langle\Delta r^2(2\pi/\omega)\rangle \Gamma[1+\alpha(\omega)],$$

where

$$\alpha(\omega) := \left.\frac{d\log\Delta r^2(\tau)}{d\log\tau}\right|_{\tau=2\pi/\omega}$$

is the local power law exponent of $\langle\Delta r^2(\tau)\rangle$ and $\Gamma(\ldots)$ is the Euler gamma function. The values of $\alpha(\omega)$ for the MSDs shown in Fig. S.7B. In general $\alpha$ varies between diffusive, $\alpha \approx 1$, and sub-diffusive $\alpha \approx 0.3$ in the range of lag times sampled here. At long times the scaling exponents are diffusive, i.e. $\alpha \approx 1$, and reflect the diffusion of the probe bead in the solvent. The minimum in $\alpha$ moves to shorter times (i.e. higher frequencies) and becomes deeper as the mesh size of the solution, $\xi/L$, is made smaller. The uncertainty in $\langle\overline{\Delta r^2}[\omega]\rangle$ is estimated by standard propagation of error as

$$\delta\langle\overline{\Delta r^2}[\omega]\rangle = \sqrt{\left[\frac{d\langle\overline{\Delta r^2}[\omega]\rangle}{d\langle\Delta r^2(\tau)\rangle}\sigma(\langle\Delta r^2(\tau)\rangle)\right]^2 + \left[\frac{d\langle\overline{\Delta r^2}[\omega]\rangle}{d\alpha}\delta\alpha\right]^2},$$

whereas pointed out before $\sigma(\langle\Delta r^2(\tau)\rangle)$ is obtained using MUnCH (20) and the uncertainty in $\alpha$ is obtained from propagation of error,

$$\delta\alpha = \sqrt{\left[\frac{d\alpha}{d\langle\Delta r^2(\tau)\rangle}\sigma(\langle\Delta r^2(\tau)\rangle)\right]^2}.$$

Using the above equation, the uncertainty in $G^*(\omega)$ is obtained by the propagation of error as,

$$\delta G^*(\omega) = \sqrt{\left[\frac{dG^*(\omega)}{d\langle\overline{\Delta r^2}[\omega]\rangle}\delta\langle\overline{\Delta r^2}[\omega]\rangle\right]^2}.$$

The storage modulus, $G'(\omega) = Re\{G^*(\omega)\}$ and its uncertainty, the loss modulus, $G''(\omega) = Im\{G^*(\omega)\}$, and its uncertainty are shown in the main text. The zero-shear viscosity is obtained from

$$\eta = \int_0^\infty G(t)\, dt,$$

where the relaxation modulus, $G(t)$, is related to dynamic modulus by

$$G^*(\omega) = i\omega \overline{\mathcal{F}}[G(t)] = i\omega \int_0^\infty G(t)\, e^{-i\omega t} dt.$$

Tables

Table. S1. Simulation parameters

| parameter | Value (simulation unit) |
|---|---|
| Thermal energy ($\varepsilon$) | $1[E]$ |
| Nanofiber width ($b$) | $1[x]$ |
| Time step ($\tau$) | $0.001[t]$ |
| LB time step ($\tau_f$) | $0.001[t]$ |
| Lattice constant ($a$) | $1[x]$ |
| Kinematic viscosity ($\nu$) | $\nu_w \cdot \dfrac{\tau_f}{a^2}$ |
| Solvent density ($\rho$) | $k_B T \cdot \dfrac{\tau_f^{\,2}}{a^5 \cdot 1\varepsilon}$ |
| $\sigma$ in WCA potential | $1[x]$ |
| $r_{cut}$ in WCA potential | $1[x] \cdot 2^{1/6}$ |
| $r_0$ in harmonic potential | $1[x]$ |
| $\gamma$ friction parameter | $10 \dfrac{[m]}{[t]}$ |
| Nanofiber length ($L$) | $50[x] \leq L \leq 300[x]$ |
| Box size (box_l) | $minimum \to 1.3\,L$ |

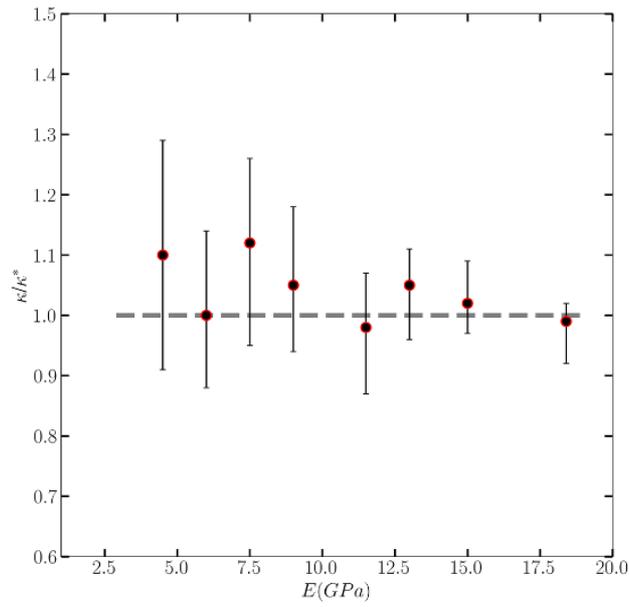

**Fig. S1**. The ratio of bending stiffness from Young's modulus to thermal fluctuation motion $\kappa/\kappa^*$. When comparing examples of different stiffnesses, it is evident that the error is bigger for less stiff cases owing to the higher rate of thermal motion.

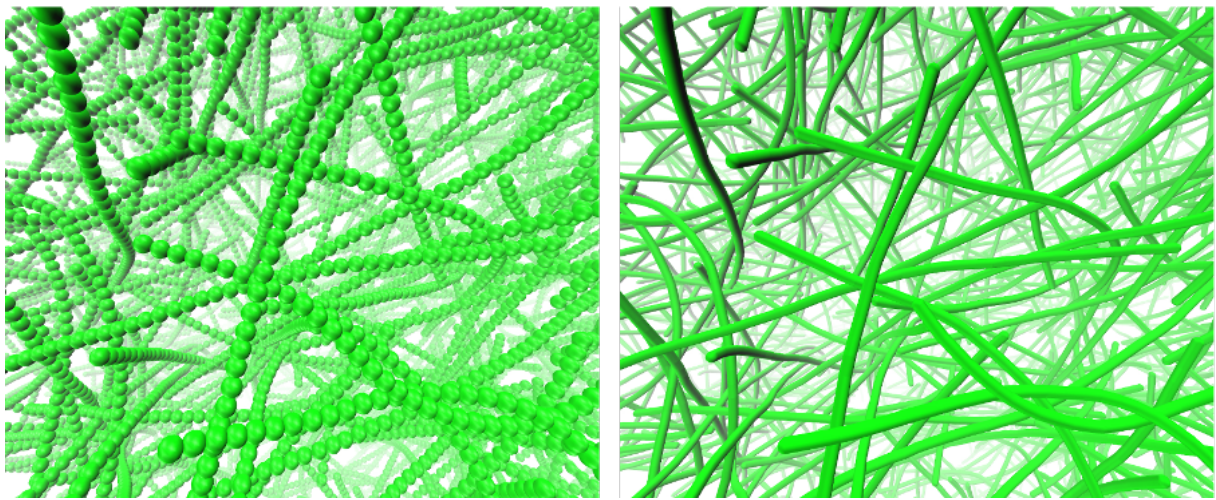

**Fig. S2**. A demonstration of nanofiber modeling for the special situation of $L_p/L=1.5$ and $\xi/L=0.12$. In supplementary Movie S1, the dynamics of nanofiber simulations are also shown.

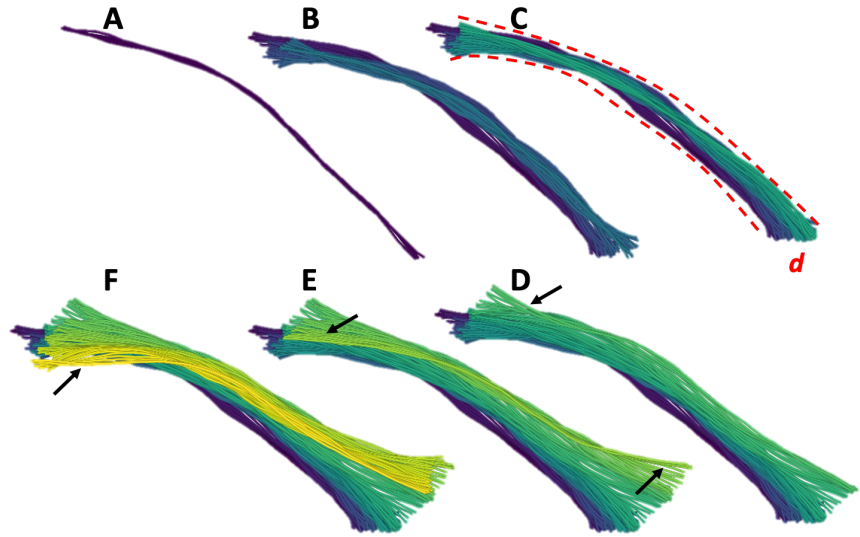

**Fig. S3.** The time-dependent behavior of a single nanofiber in the entangled system of $L_p/L=7.2$ and $\xi_p/L=0.098$. The color map shows relative timing. **(A)** shows the first stage of the motion. (B) The nanofiber reptates along its length in the tube. (C) At this moment, the tube size and shape are distinguishable. (D) The nanofiber is liberated from the tube and the constraint is released. (E) After being liberated, the nanofiber explores further and shows a turning move, coupling translation and rotation. (F) The nanofiber continues to diffuse freely in space until being trapped in a new tube.

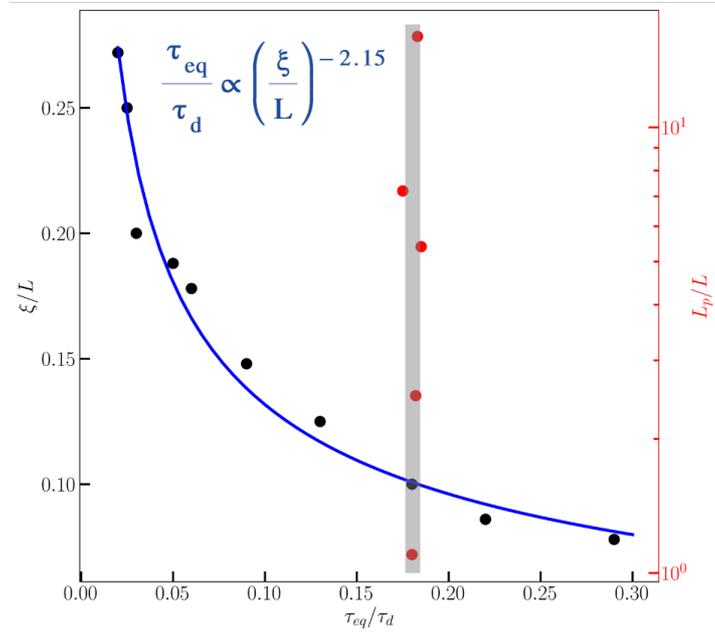

**Fig. S4.** The dependency of $\tau_{eq}$ on system crowding ($\xi_p/L$) and nanofiber stiffness ($L_p/L$) is shown. The black circles depict the relationship of $\tau_{eq}$ to $\xi_p/L$, whereas the red circles depict its relationship to $L_p/L$ for special case of $\xi_p/L=0.11$. Internal relaxation time clearly depends heavily on $\xi_p/L$ as $\left(\tau_{eq}/\tau_d\right) \propto \left(\xi_p/L\right)^{-2.15}$. In a sense, it takes more time for internal relaxations to take place in a crowded environment than in a less crowded one. The internal relaxation time, however, is independent of nanofiber stiffness.

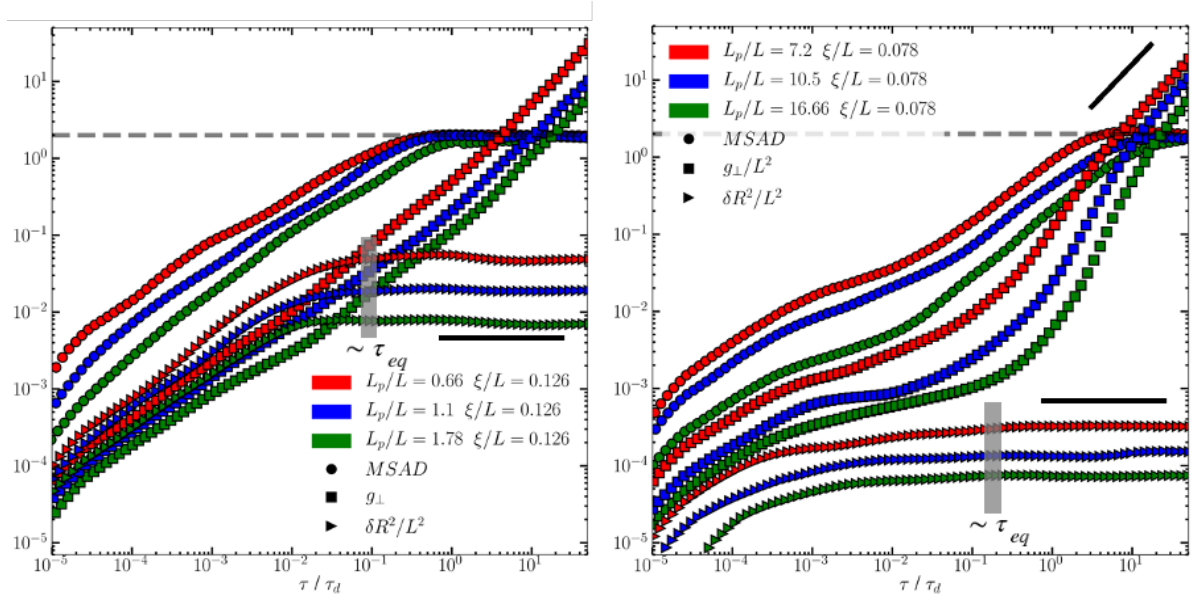

**Fig. S5.** Plotted observables for various cases. As is evident from the $\delta R^2/L^2$ behavior, nanofibers need more time to relax internally in more crowded surroundings (lower $\xi_p/L$). In addition, based on $g_\perp/L^2$ and $MSAD$, we may conclude that confinement effects are less observable in less crowded environments.

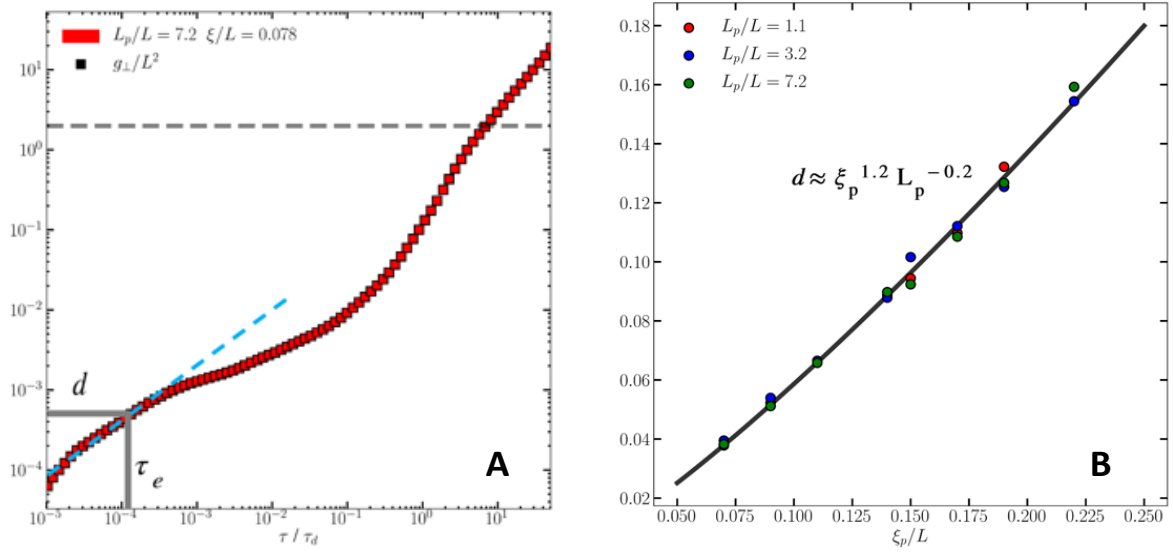

**Fig. S6.** (A) $g_\perp(t)$ plot for the case with $L_p/L=7.2$ and $\xi_p/L=0.078$. $g_\perp(t)$ shows the identical behavior of $\delta R^2(t)$ for short times as $t^{3/4}$. The tube diameter is roughly estimated based on the transition point from $t^{3/4}$ behavior to the entangled regime. An approximation of the entanglement time, $\tau_e$, may also be obtained by using the point's time coordinate. (B) The correlation between the predicted effective tube diameter $d$, $\xi_p$ and $L_p$ of the system as $d \approx \xi_p^{1.2} L_p^{-0.2}$. Dissimilar to the conventional assumption that holds that the tube diameter is the same as the mesh size, the tube diameter is changing with the mesh size.

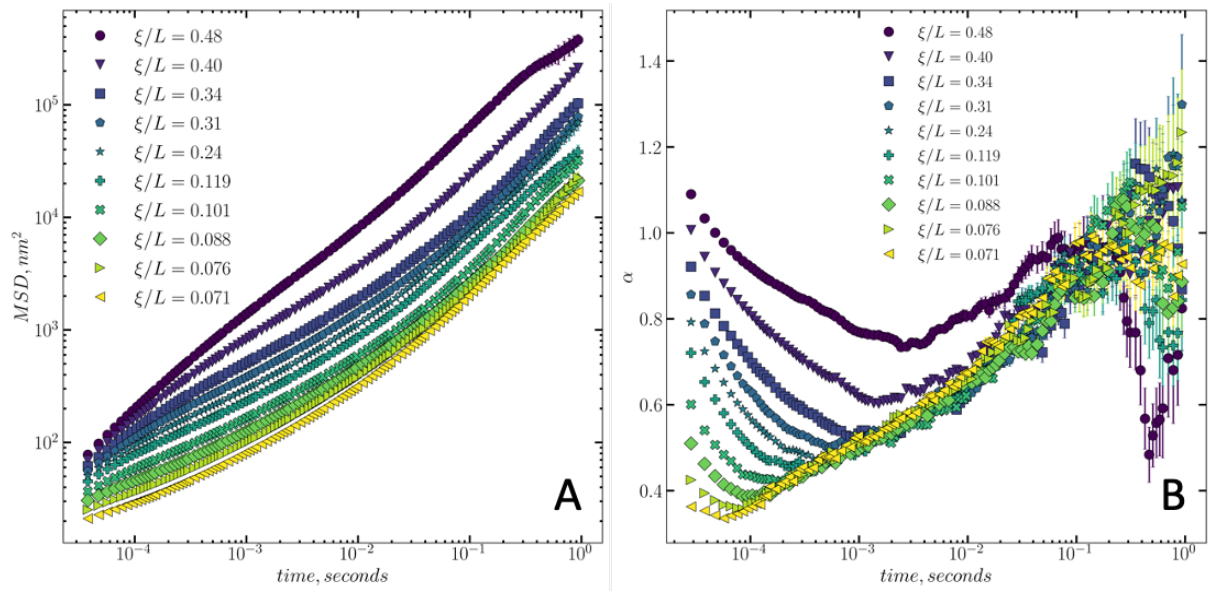

**Fig. S7.** A) Mean-squared displacement of the probe bead as a function of lag time in CNF networks with different mesh sizes, $\xi_p/L$. The MSD and its error bars are calculated using MUnCH(20, 21). B) Local exponent in the MSD, $\alpha$ as a function of lag time for the MSDs shown in part A).

**Movies**

Movie S1. A demonstration of nanofiber modeling for the special situation of $L_p/L = 1.5$ and $\xi/L = 0.12$

Description: 3D rotating view of a simulated polymer network using bead-spring model, highlighting the structure and dynamics of the simulated polymer systems.

Movie S2. Dynamics of a single nanofiber for the case of $L_p/L = 7.2$ and $\xi_p/L = 0.098$: breaking the cage

Description: The movie demonstrates the dynamics of a semi-flexible nanofiber moving in a crowded environment of other dynamic polymers. The nanofiber initially is trapped in a tube but eventually breaks free and continues to rotate and find the next tube. This movie depicts the constraint release mechanism and presence of dynamic cages.

Movie S3. Dynamics of a single fiber: breaking the cage

Description: The movie demonstrates the dynamics of a semi-flexible nanofiber moving in a crowded environment of other dynamic polymers. This movie depicts the constraint release mechanism and presence of dynamic cages for a shorter nanofiber compared to Movie S2.

Movie S4. Dynamics of a single nanofiber for the case of $L_p/L = 1.9$ and $\xi_p/L = 0.14$ (stiffness effect)

Description: Simulated dynamics of highly flexible nanofibers in a dynamic environment, highlighting early entanglement compared to networks made of nanofibers that are relatively less flexible.

Movie S5. Dynamics of a single nanofiber for the case of $L_p/L = 3.6$ and $\xi_p/L = 0.14$ (stiffness effect)

Description: Simulated network of nanofibers with lower flexibility, exhibiting lower degree of entanglement compared to a more flexible network.

Movie S6. A test rigid polymer in a fixed static network

Description: The movie presents dynamics of a rigid rod in a static network of other rigid rods. The simulations demonstrate the stability of tubes or cages within the network, providing insight into the behavior of a stiff polymer in a stable tube.

Movie S7. A test rigid polymer in a dynamic system (Kinetic cages)

Description: The movie shows dynamics of a rigid rod in a dynamic network of other moving rods. The dynamics reveal the kinetic nature of the tubes or cages within the network, which have a short lifetime compared to those in a static network.

Movie S8. Empty spaces between the nanofibers with time

Description: The movie aims to provide insight into the creation and dissolution of tubes or cages within a dynamic network of nanofibers. The simulations showcase the empty spaces between the nanofibers and how they contribute to the formation and breakdown of these structures over time.